\documentstyle[preprint,aps]{revtex}
\begin{document}
\draft

\title{STABILITY OF PERIODIC ARRAYS OF VORTICES}
\author{Thierry DAUXOIS\thanks{tdauxois@physique.ens-lyon.fr},
 Stephan FAUVE and Laurette TUCKERMAN\thanks{
Permanent address: LIMSI, BP 133, 91403 Orsay Cedex, France}}
\address{Laboratoire de Physique, URA-CNRS No. 1325,
Ecole Normale Sup\'erieure de Lyon, \\
46 All\'{e}e d'Italie, 69364 Lyon Cedex 07, France}
\date{\today}
\maketitle
\begin{abstract}
The stability of periodic arrays of Mallier-Maslowe or
Kelvin-Stuart vortices  is discussed. We derive with the
energy-Casimir stability method the nonlinear  stability of
this solution in the inviscid case as a function of the
solution  parameters and of the domain size. We exhibit the
maximum size of the domain  for which the vortex street is
stable.

By adapting a numerical time-stepping code, we calculate the
linear stability  of the Mallier-Maslowe solution in the
presence of viscosity and compensating  forcing. Finally,
the results are discussed and compared to a recent
experiment  in fluids performed by Tabeling et
al.~[Europhysics Letters {\bf 3}, 459 (1987)].
Electromagnetically driven counter-rotating vortices are
unstable above a  critical electric current, and give way to
co-rotating vortices. The importance  of the friction at the
bottom  of the experimental apparatus is also discussed.
\end{abstract}
\vskip 1truecm
\vfill\eject

\section{Introduction}

The problem of vortex dynamics is important for the field of
chaotic motion and dynamical systems theory, but the
discovery of coherent structures in turbulence has fostered
the hope that the study of vortices will also lead to
 a better  understanding of turbulent
flows~\cite{SAFFMAN}. The emergence of coherent flow
structures is a well-known feature of quasi-geostrophic
flows~\cite{FLIERL}, soap films or two-dimensional
turbulence\cite{COUDERTBASDEVANT} and,  because of their
relevance to large-scale geophysical flows, the dynamics of
these structures has attracted attention during the past two
decades. Geophysical fluid flows often appear to be
dominated by  a strong but localized vortical structure that
lasts for many circulation times even when relatively
turbulent flows are impinging upon it. Experimental evidence
indicates also that the planar free  shear layer has an
organized two-dimensional structure over a wide range of
Reynolds numbers~\cite{PIERREHUMBERWIDNAL,HEIJSTFLOR}.  When
modelling steady-state configurations of geophysical flows
as solutions of a dynamical system, it is
important to analyze their stability  in
order to see if they can describe physically observable
situations. Indeed, in the real world, many random forces
act on the system and the stationary  situations we observe
must be stable under these perturbations.

The dynamics of coherent structures in two-dimensional
geometry has been studied in  many different experiments
using rotating or stratified  fluid
(see~\cite{HOPFINGERVANHEIJST} and references therein), in a
shallow layer of mercury~\cite{NGUYENDUCSOMMERIA}  or of
electrolyte~\cite{BONDARENKOGAK} subjected to a magnetic
field. Here let us recall the experiment proposed by
Tabeling et al~\cite{TABELINGPERRINFAUVE}.  A periodic array
of counter-rotating vortices is driven by electromagnetic
forcing. By passing a current through  a cell containing a
solution  of sulfuric acid and an array of permanent magnets
of alternating polarity at the bottom of the cell, the
Lorentz force stirs the fluid, producing the vortices. The
two-dimensionality of the flow is ensured using a shallow
fluid layer. The basic results can be summarized as follows:

(i) At low current corresponding to weak forcing and hence
low Reynolds number, the flow consists of a linear array of
counter-rotating vortices.

(ii) This state becomes unstable beyond a critical current.
The linear array is now composed of nonuniform tilted
vortices, alternately large and small.

(iii) A further increase in the current leads to a state with
half the number of co-rotating vortices as compared to the
initial state.

This experiment has led to a number of studies concerning chaotic
regimes\cite{CARDOSMARTEAUTABELING}. Let us present another
point of view. The first question we want to address is the
following: what is the connection between the patterns of
the Navier-Stokes equation and the exact solutions of the
Euler equation, where solutions of this type are known to
exist~? The second question is to determine the stability of
such coherent structures in the presence of viscosity and
forcing.

We have organized the article in the following way. In
section~\ref{INVISCIDFLOW}, we review some steady-state
solutions of the two-dimensional inviscid and incompressible
fluid motion. We will also present
the Mallier-Maslowe vortex street that we will study
in the remainder of the article.  Sections~\ref{nonlinearstability},
{}~\ref{VISCOUSANDFORCEDFLOW} and~\ref{timeintegration} form
the heart of the paper.  In section~\ref{nonlinearstability}, we
derive analytically  explicit sufficient conditions for the
nonlinear stability estimates in the inviscid case, using
Casimirs and convexity properties.  In
section~\ref{VISCOUSANDFORCEDFLOW},  we discuss the
two-dimensional viscous flows and present the numerical
method used for studying the linear stability.
Section~\ref{linearstability} discusses the results.
Section~\ref{Discussion} sets up the correspondence  between
the results and the experiment. Finally, in section
\ref{timeintegration}, the nonlinear evolution of an unstable
Mallier-Maslowe solution is
presented. The results are then discussed in connection with
the experiments of Tabeling et
al~\cite{TABELINGPERRINFAUVE}.

\section{INVISCID FLOW}
\label{INVISCIDFLOW}

For two-dimensional incompressible fluid motion, one
obtains from the Navier-Stokes equation, by elimination of
the pressure, the  equation for the stream function~$\psi$:
\begin{equation}
{\partial\nabla^2\psi\over \partial t}+J(\nabla^2\psi,\psi)
=\nu\nabla^4\psi+G_{ext}
\label{eq:equationdeNS}
\end{equation}
where $J(A,B)=A_xB_y-A_yB_x$ is the usual Poisson bracket and
$\nu$ the kinematic viscosity. The last term  $G_{ext}$ is
due to possible external forcing.

In the absence of external forcing and viscosity,
 the vorticity equation  reduces to
the Euler equation:
\begin{equation}
{\partial\nabla^2\psi\over \partial t}=J(\psi,\nabla^2\psi)
\label{eq:equationenpsi}
\end{equation}
For steady-state flows, this gives
\begin{equation}
J(\psi,\nabla^2\psi)=0\quad .
\label{stationaireEuler}
\end{equation}
 Hence it
follows that the vorticity $\omega=-\nabla^2\psi$ is
constant along contours of constant stream function~$\psi$.
The study of planar steady-state flows in an ideal
incompressible liquid is consequently reduced  to solving
the following equation:
\begin{equation}
\nabla^2\psi
={\partial^2\psi\over \partial x^2}+ {\partial^2\psi\over \partial
y^2}=F(\psi)
\label{followingequ}
\end{equation}
where $F$ is an arbitrary function. This is a nonlinear
elliptic equation for~$\psi$ and therefore
admits a continuous multiplicity of solutions
associated with the arbitrariness of $F(\psi)$.
The problem
of finding steady states of two-dimensional vortices in an
inviscid fluid is then equivalent to solving the Poisson
equation for the electrostatic potential with the charge
density self-consistently determined.

The simplest choice for $F$ is a linear function,
which already  gives many different patterns. Indeed
Kolmogorov flows, cellular structures with square  or
hexagonal cells and even quasi-crystal patterns are
solutions\cite{ZASLAVSKYETAL} of the Helmholtz equation
$\nabla^2\psi =-\psi$. Many other solutions have been
studied in the literature such as the Lamb
dipole~\cite{LAMB,RASMUSSEN} and the non-symmetric Chaplygin
dipolar solutions~\cite{MELESHKOHEIJST}.

A possible choice for the function $F$
which has been proposed in the literature
\cite{TINGCHENLEEa,TINGCHENLEEb,PASMANTER} is
 \begin{equation}
{\partial^2\psi\over \partial x^2}+ {\partial^2\psi\over
\partial y^2}=-{(1-\varepsilon^2)\over2} \sinh (2\psi)\quad.
\label{sinhPoisson}
\end{equation}
This $\omega-\psi$ sinh-relationship is very important because,
using a statistical approach, one can
show~\cite{MONGOMMERYJYCE}  that it characterizes the most
probable state of a two-dimensional system of ideal point
vortices. Published data showing the functional dependence
of vorticity on stream function in long-lived structures,
seen in experiments and simulations, seem qualitatively
consistent with hyperbolic-sine  (as in
Eq.~\ref{sinhPoisson}) or exponential profiles (as in
Eq.~\ref{equationdeStuart} below)  which follow from entropy
maximization~\cite{SOMMERIA}. These studies have been
verified by very long-time  high resolution numerical
studies; however, recent work shows some evidence that
this result depends on the initial
conditions~\cite{RASMUSSEN}. Another recent experimental
study~\cite{MARTEAUCATTAB} led to the conclusion that the maximum
entropy state is unlikely to be reached since the observed
final states of flow display characteristics that conflict
with the statistical theory.

In the case of an infinite box,  the solution to
Eq.~(\ref{sinhPoisson}) introduced by Mallier and
Maslowe~\cite{MALLIERMASLOWE} which we will discuss in the remainder
of the article is:
\begin{equation}
\psi_M=\ln\Biggl({\cosh{ \varepsilon y}-
\varepsilon\cos x\over\cosh{ \varepsilon y}+ \varepsilon\cos
x} \Biggr) =-2\ \hbox{\rm arctanh}\Biggl({  \varepsilon\cos
x\over \cosh{ \varepsilon y}}\Biggr)
\label{eq:solutionalternee}
\end{equation}
which describes a stationary  pattern in the form of a
street of counter-rotating vortices, arranged periodically
along the x-axis at intervals of $\pi$. A typical
 solution is shown in  figure~\ref{profileofthesolution}(a)
for $\varepsilon=0.3$.  The
parameter $\varepsilon$ characterizes  the vorticity
density: when $\varepsilon=\pm 1$, we recover the point
vortex solution and when $\varepsilon=0$ we have $\psi=0$.
Thus, as $\varepsilon$ ranges from 0 to 1, the flow
represented by Eq.~(\ref{eq:solutionalternee}) ranges from
the fluid  at rest, to the flow due to a set  of point
vortices on the $x$-axis.

A third choice for the function $F$
of Eq.~(\ref{followingequ}) is that of J. T.
Stuart~\cite{STUART}, an
exponential function:
\begin{equation}
{\partial^2\psi\over \partial x^2}+ {\partial^2\psi\over
\partial y^2}=(1-\varepsilon^2)e^{-2\psi} \quad .
\label{equationdeStuart}
\end{equation}
If we use the change of variable proposed above, this equation
is directly related to the well-known Liouville equation for
a real scalar field $\psi(y,t)$, studied by both
Liouville~\cite{LIOUVILLE} and by Poincar\'e~\cite{POINCARE}.

The exact nonlinear
solution to Eq.~(\ref{equationdeStuart}):
\begin{equation}
\psi_S=\ln\left({\cosh{y}+ \varepsilon\cos x}\right)\quad .
\label{eq:solutionStuart}
\end{equation}
is called Kelvin-Stuart's cat's eyes \cite{STUART}
and is illustrated in Fig.~\ref{profileofthesolution}(b).
This solution can be derived in an elegant way with the
Hirota method, assuming a vortex spacing of $2\pi$.
This solution is of interest because the solution
corresponds qualitatively to the co-rotating vortices seen by
Tabeling et al.~\cite{TABELINGPERRINFAUVE}.

Analytical expressions of the co-rotating or counter-rotating
vortex streets are especially useful for studying the stability
of the experimental fluid flows~\cite{TABELINGPERRINFAUVE}
described before.  It would be of interest to
find an analytic function connecting the
co-rotating and counter-rotating solutions.
As written by Mallier and
Maslowe~\cite{MALLIERMASLOWE},
 if such a function exists, it is likely that at
the best the general equation will be the following:
\begin{equation}
{\partial^2\psi\over \partial x^2}+ {\partial^2\psi\over
\partial y^2}=Ae^{2\psi}+Be^{-2\psi} \quad .
\label{Stuart+mallier}
\end{equation}
It can be shown that, rather than interpolating between
Eq.~(\ref{sinhPoisson}) and Eq.~(\ref{equationdeStuart}),
equation~(\ref{Stuart+mallier}) reduces to
the sinh-Poisson equation~(\ref{sinhPoisson}) for all nonzero values of $A$.
Therefore a solution of equation~(\ref{Stuart+mallier}) is
\begin{equation}
\psi=\ln\Biggl({\cosh{ \varepsilon y}-
\varepsilon\cos x\over\cosh{ \varepsilon y}+ \varepsilon\cos
x} \Biggr)  -\psi_0\quad ,
\label{solformel}
\end{equation}
where  $\varepsilon=\left(1-4\sqrt{-AB}
\right)^{-\slantfrac{1}{2}}$ and
$\psi_0={1\over2}\ {\rm arccosh} \left[(B-A)/
2(-AB)^{\slantfrac{1}{2}}\right]$.

Moreover, it is possible to treat
the  Liouville equation~(\ref{equationdeStuart})
corresponding to the limiting case $A=0$,
as a singular limit of the  sinh-Poisson
equation~\cite{GERVAISNEVEU} by making the substitution
$1-\varepsilon^2=\lambda^2e^{-2\beta}$ and
$\psi=u-\beta$ in equation~(\ref{sinhPoisson}),
 and then taking $\beta\rightarrow+\infty$.
By carefully following what happens in this highly singular
limit, Tracy et al~\cite{TRACYCHINCHEN} succeeded in
exhibiting the Liouville solution as a singular limit of the
sinh-Poisson solution. However, to our knowledge, the
function connecting the  two solutions has not been found.
At this point, let us turn our attention to determining the
nonlinear stability of the Mallier-Maslowe solutions.

\section{Nonlinear stability of the counter-rotating vortices}
\label{nonlinearstability}
We are interested in the stability of
the Mallier-Maslowe  solution~(\ref{eq:solutionalternee}) in
order to explain the experimental results presented
by Tabeling et al.~\cite{TABELINGPERRINFAUVE}.
To establish explicit sufficient stability
conditions~\cite{HOLMLAM} for all values of $\varepsilon$
and to study the nonlinear stability of the counter-rotating
vortices in a domain $D$ of the plane R$^2$, in the former
article~\cite{DAUXOIS}, we used the total energy on this
domain
\begin{eqnarray}
H(\omega)&=&\int\!\!\!\int_D^{}{1\over
2}|\overrightarrow{v}|^2\ dx\ dy\nonumber\\ &=&
\int_{\partial D}^{}{1\over 2}
\psi\overrightarrow{\nabla}\psi .\overrightarrow{n} ds
-\int\!\!\!\int_D^{}{1\over 2}\psi\nabla^2\psi\ dx\
dy\nonumber\\ &=& {1\over 2}\int\!\!\!\int_D^{}\psi\omega\
dx\ dy
\end{eqnarray}
where we have used the fact that the velocity,
and hence $\overrightarrow{\nabla}\psi$,
vanishes on the boundary of $D$.
Since the fluid is inviscid, this
quantity is conserved. More generally, one can also
show~\cite{HOLMPR}, that the functionals  $\displaystyle
C_\Phi(\omega)=\int\!\!\!\int_D^{}\Phi(\omega)\ dx\ dy$, called
Casimirs, are also conserved for any real-valued function $\Phi$.

We define a conserved quantity $H_\Phi\equiv H+C_\Phi$
whose functional derivative is:
\begin{eqnarray}
DH_\Phi(\omega).\delta\omega
&=&\int\!\!\!\int_{D}^{}\left(\psi(\omega)
+\Phi^{'}(\omega)\right)\delta\omega\
dx\ dy \quad.
\end{eqnarray}
We wish to choose  $\Phi$
so that $DH_\Phi(\omega_M)=0$,
where $\omega_M=-\nabla^2\psi_M$ and $\psi_M$
is defined by Eq.~(\ref{eq:solutionalternee}).
We obtain
\begin{equation}
-\Phi^{''}(\omega) = \psi_M^{'}(\omega) = {1\over
\sqrt{4\omega^2+(1- \varepsilon^2)^2}}\quad .
\label{derive}
\end{equation}
leading to
$\left(1-\varepsilon^2\right)^{-1}
\geq -\Phi^{''}(\omega) \geq 0$.
We will need to bound $-\Phi^{''}$ away from zero.
Eq.~(\ref{sinhPoisson}-\ref{eq:solutionalternee}) state that
\begin{equation}
\omega = {{1-\varepsilon^2}\over 2} \sinh(4\hbox{\rm Arcth}\ g(x,y))
\hbox{\rm ~~~~~~with~~~~}
|g(x,y)| = \left|{{\varepsilon \cos x}\over {\cosh \varepsilon y}}\right|
\leq \varepsilon
\end{equation}
so that $\omega$ is bounded by
$|\omega| \leq \omega_{max} \equiv {{1-\varepsilon^2}\over 2}
\sinh(4\hbox{\rm Arcth}\ \varepsilon)$.
The calculation
\begin{equation}
\sqrt{4\omega^2+(1- \varepsilon^2)^2} \leq (1-\varepsilon^2)
\cosh(4\hbox{\rm Arcth}\ \varepsilon ) =
{{1+6 \varepsilon^2+ \varepsilon^4}\over{1-\varepsilon^2}}
\end{equation}
leads to the improved bounds
\begin{equation}
{{-1}\over{1-\varepsilon^2}}\leq \Phi^{''}(\omega) \leq
{{-(1-\varepsilon^2)}\over {1+6 \varepsilon^2+ \varepsilon^4}}\quad.
\label{betterbounds}
\end{equation}

However, the bounds in (\ref{betterbounds}) apply only to
$|\omega| \leq \omega_{\max}$,
whereas we will require such bounds to hold over the
entire real line.
We therefore construct  a function ${\tilde{\Phi}}$
to coincide  with $\Phi$ for $|\omega| \leq \omega_{\max}$ and with
\begin{equation}
\tilde{\Phi}(\omega)=-\Biggl({1-\varepsilon^2\over
1+6 \varepsilon^2+ \varepsilon^4}\Biggr){\omega^2\over 2}
 +\alpha_{\pm}\omega+\beta_{\pm}
\end{equation}
for $|\omega| \geq \omega_{\max}$.
The constants
 $\alpha_{\pm}$ and $\beta_{\pm}$
 are determined by continuity,
 so that $\tilde{\Phi}$ is a C$^2$-function.

With these preparations completed, we are ready to define
the nonlinear constant of motion:
\begin{eqnarray}
\hat H_{\tilde\Phi}(\delta\omega)
&\equiv&
H_{\tilde\Phi}(\omega_M+\delta\omega)-H_{\tilde\Phi}(\omega_M)-
DH_{\tilde\Phi}(\omega_M).\delta \omega\nonumber\\
&=&\int\!\!\!\int_{D}^{}\left[{1\over 2}\delta \omega\ (-\nabla^2)^{-1}
\delta \omega+
{\tilde{\Phi}}(\omega_M+\delta\omega)-{\tilde{\Phi}}(\omega_M)-
{\tilde{\Phi}}^{'}(\omega_M).\delta \omega\right]
\ dx \ dy
\label{Hchapeau}\end{eqnarray}
and to use it to establish Liapunov stability estimates.
Using the bounds (\ref{betterbounds}), we get,
\begin{equation}
\left({1-\varepsilon^2\over
1+6 \varepsilon^2+ \varepsilon^4}\right){\delta\omega^2\over 2}
\leq -{\tilde{\Phi}}(\omega_M+\delta\omega)+{\tilde{\Phi}}(\omega_M)+
{\tilde{\Phi}}^{'}(\omega_M).\delta \omega
\leq {1\over1- \varepsilon^2}{\delta\omega^2\over 2}\quad .
\label{agarder}
\end{equation}
We introduce $k_{\min}^2$, the minimal  eigenvalue of the
positive operator $(-\bigtriangledown^2)$, to obtain
\begin{equation}
0\leq
\int\!\!\!\int_{D}^{}{1\over 2}\delta \omega\ (-\nabla^2)^{-1}
\delta \omega
\leq {1\over 2} k_{\min}^{-2} ||\delta \omega||_{L^2}^2 \quad.
\label{bgarder}
\end{equation}
Combining (\ref{agarder}) and (\ref{bgarder}), we have
\begin{equation}
\Biggl({{1-\varepsilon^2}\over{1+6\varepsilon^2+\varepsilon^4}}
-k_{\min}^{-2}\Biggr) ||\delta \omega||_{L^2}^2
\leq
-2{\hat{H}}_{  {\tilde{\Phi}}}(\delta\omega)
\leq
{1\over{1-\varepsilon^2}}||\delta \omega||_{L^2}^2 \quad.
\end{equation}

Now consider an initial value of the perturbation $\delta\omega_0$.
Since ${\hat{H}}_{{\tilde{\Phi}}}$ is a conserved quantity,
\begin{equation}
-2{\hat{H}}_{{\tilde{\Phi}}}(\delta\omega)
=-2{\hat{H}}_{  {\tilde{\Phi}}}(\delta \omega_0)
\leq {1\over {1- \varepsilon^2}} ||\delta \omega_0||_{L^2}^2
\end{equation}
This {\sl a priori} estimate provides suitable norms bounding
the growth of disturbances
since we have finally
\begin{equation}
\Biggl[ {1- \varepsilon^2\over
1+6 \varepsilon^2+ \varepsilon^4}
 -k_{\min}^{-2}\Biggr]||\delta \omega||_{L^2}^2\leq
{1\over(1- \varepsilon^2)}\ ||\delta \omega_0||_{L^2}^2
\quad .
\label{eq:formulestabilitefinale}
\end{equation}
The solution is nonlinearly stable if the term in brackets is
positive.

Consider for the domain $D$ a rectangular box, with length
$2\pi N$ in $x$ and $2\ell$ in $y$; the
minimal eigenvalue of the operator  $(-\bigtriangledown^2)$
is   $k_{\min}^2={(1/N^2)}+{(\pi^2 /\ell^2)}$, since the
eigenfunctions vanishing on the boundary are $f(x,y)=\cos
{(x/ N)}\sin {(\pi y/ \ell)}$.
Therefore, we have derived a maximum transverse size of the domain D
for which the Mallier-Maslowe vortex street is nonlinearly
stable.
The sufficient conditional stability is the following:
 \begin{equation}
{\pi\over \ell}> \sqrt{{1+6 \varepsilon^2+
\varepsilon^4\over1- \varepsilon^2}-{1\over N^2}}\; .
\label{eq:conditionstab}
\end{equation}
Figure~\ref{lfonctiondeepsilon} presents the
region of sufficient stability of the counter-rotating vortices in the
($\ell,\varepsilon$) plane for $N=1$.

\section{VISCOUS AND FORCED FLOW}
\label{VISCOUSANDFORCEDFLOW}

\subsection{Introduction}

It would be interesting to extend the previously presented
results in the presence of viscosity and forcing. Moreover
 the viscosity imposes a minimum
 scale $(\nu/\sup\
\omega)^{1\over 2}$, the diffusion length for the eddy
turn-over period at the maximum realised vorticity. Thus, a small
viscosity avoids some difficulties concerning the continuum
limit of Euler flow. To fully understand the nonlinear
evolution, we can follow the time evolution of the system from
various initial conditions; however, it is useful to obtain
the eigenspectrum of the steady states, since they are
associated with  transitions and loss of stability.

In the viscous case, we have the full two-dimensional
Navier-Stokes equation
\begin{equation}
{\partial\nabla^2\psi\over \partial t}+J(\nabla^2\psi,\psi)
=\nu\nabla^4\psi+G_{ext}
\label{eq:equationdeNScomplete}
\end{equation}
In what follows, we will
 choose the external forcing to counterbalance the viscosity:
\begin{equation}
G_{ext}=-\nu\nabla^4\psi_M
\label{forceext}
\end{equation}
for which we shall give a partial justification
in section~\ref{Discussion}. With the
choice~(\ref{forceext}), if $\psi_M$ is a stationary solution
to the Euler equation~(\ref{stationaireEuler}),
then $\psi_M$ is also a solution to the the full Navier-Stokes
equation~(\ref{eq:equationdeNScomplete}).
Viscosity plays an important role, however,
 in determining the {\it stability} of the solution $\psi_M$.
Our
strategy is to linearize the Navier-Stokes equations around
the steady states and to seek eigenmodes of the linearized
equations. The linearized equation governing the evolution of a
perturbation $\phi$ is
\begin{equation}
{\partial  \nabla^2\phi\over \partial t}=J(\psi_M, \nabla^2\phi)+
J(\phi,\nabla^2\psi_M)+\nu\nabla^4\phi\quad .
\label{eq:equationenphi}
\end{equation}

\subsection{Numerical procedure}
\label{Numericalprocedure}

All of our calculations are performed on a two-dimensional $(x,y)$ plane.
In the periodic $x$-direction we use
a Fourier representation with $N_x$ modes (from 16 to 64);
We map $y \in (-\infty,+\infty)$ to $(-1,+1)$ via a tanh mapping
 with $N_y$ gridpoints
(from 65 to 123).
Boundary conditions are automatically satisfied in this
representation: $\phi(x+2\pi,y)=\phi(x,y)$ and
$\partial \phi/\partial y (x,y=\pm\infty) = 0$.

For stability, the viscous term
$\nabla^4\phi$ in Eq.~(\ref{eq:equationenphi})
is integrated implicitly by the backward Euler scheme. The
remaining terms are integrated explicitly.
We have
\begin{equation}
\nabla^2 \phi_{n+1}=(I-\nu \Delta t\nabla^2)^{-1}\ \left[
\nabla^2\phi_n+ \Delta t\left(J(\psi_M,\nabla^2\phi_n)+
J(\phi_n,\nabla^2\psi_M)\right)\right]\quad .
\label{addi}
\end{equation}
where $\Delta t$ is the time step.

The linear stability of $\psi_M$ is determined by the leading
eigenvalues (those with greatest real part) of the
operator
on the right hand side of~(\ref{eq:equationenphi}). The
leading eigenvalues of the operator of the differential
equation~(\ref{eq:equationenphi}) become the dominant ones
(those with largest magnitude) of the iterative scheme
of Eq.~(\ref{addi}); fortunately, dominant eigenvalues
are those most readily calculated by iterative methods.
Effectively, exact solution
 of Eq.~(\ref{eq:equationenphi}) would
require  exponentiating the operator on its right hand side,
and the numerical method~(\ref{addi}) carries out an
approximate exponential.

This can be abbreviated as
\begin{equation}
{d\nabla^2 \phi\over d t}=A\ \nabla^2 \phi\quad \Leftrightarrow
 \quad \nabla^2 \phi(t)=e^{At}\ \nabla^2 \phi(0)=e^{An \Delta t}\
\nabla^2 \phi(0)=B^n \ \nabla^2 \phi(0)
\label{eq:equationenA}
\end{equation}
where $t=n  \Delta t$ and $B=e^{A \Delta t}$
is approximated by
the operator on the right-hand-side of Eq.~(\ref{addi}).

The block power, or Arnoldi's, method is used in order to
find the $k$ leading eigenvalues, including complex or
multiple eigenvalues, simultaneously, as described by Mamun
et Tuckerman~\cite{MAMUMTUCKER} and references therein.
We first  integrate Eq.~(\ref{eq:equationenphi}) for
some fairly long period of time
$T$ in order to purge the vector of the strongly damped
eigenmodes which are not important for the linear stability
study. We then take $k$ additional time steps, creating
$u_1=u(T),\dots,u_{k+1}=u(T+k dt)$. The vectors  are
orthonormalized, forming a basis for what is called  the
Krylov space. A ($k$ by $k$) matrix $H$, which represents the
action of $B$ on the  Krylov space, is generated and
diagonalized, yielding eigenvalues and
eigenvectors  of the linear stability problem. The eigenvalues
$\lambda$ of $A$ are recovered from those of $B$ (or $H$) by taking
their logarithm and dividing by $\Delta t$.

\subsection{Results of the linear stability analysis}
\label{linearstability}

Let us now study the linear stability
of the Mallier-Maslowe vortices i.e., the stability of
the flow $\psi_M$ defined by Eq.~(\ref{eq:solutionalternee})
with corresponding forcing~(\ref{forceext}).
Figure~\ref{2valpour0.3} presents the real part of the two
first eigenvalues
obtained with the above method as a function of the kinematic
viscosity when the vorticity parameter $\varepsilon$ is 0.3
in Eq.~(\ref{sinhPoisson}).

In the high viscosity regime ($\nu \gg 1$), the Jacobian
terms can be neglected and
equation~(\ref{eq:equationenphi}) becomes
\begin{equation}
{\partial\nabla^2\phi\over\partial t}=\nu\nabla^4\phi\
\label{eqdiffusion}
\end{equation}
i.e., a heat
equation for the vorticity $\nabla^2\phi$ at
$\nu\rightarrow+\infty$. In this limit, the
equation is independent of the vorticity density
parameter~$\varepsilon$, so the eigenvalues are also
independent of~$\varepsilon$. The numerical results
 confirms that, at  sufficiently large viscosity
and with the forcing chosen according to
Eq.~(\ref{forceext}), the Mallier-Maslowe
solutions~(\ref{eq:solutionalternee})
are stable, since the
growth rate of perturbations is negative.
Around the value of $\nu=0.5$, the
flow becomes unstable. The growth rate increases as the viscosity
decreases. The numerical method presented in
section~\ref{Numericalprocedure} is feasible only for high to
moderate viscosities since for low viscosities,
stability of the explicit part of the
numerical scheme requires a very
small time step, leading to a time-consuming code.
However, in the zero viscosity limit, we showed in
section~\ref{nonlinearstability} that the instability
increases with $\varepsilon$.

In order to more clearly understand the evolution of the
most unstable eigenmodes as a function of the viscosity, we will
study the two particular cases depicted by filled squares in
Fig.~\ref{2valpour0.3}, one stable ($\nu=10$) and the other
unstable  ($\nu=0.01$). Figure~\ref{Eigenvector}(a) depicts
the least stable eigenvector for $\varepsilon=0.3$ and
$\nu=10$. As we see, the eigenvector is reflection-symmetric
in $y$  and independent of $x$.  This mode resembles a shear
layer. The evolution of the eigenvalue shown in
Fig.~\ref{2valpour0.3} attests that the  $x$-independent
mode  is not the least unstable mode for low viscosity;
however, it is important for the following discussion to
notice that this mode is marginally stable for low values
of $\nu$. Figure~\ref{Eigenvector}(b) presents the contour
plot of the most unstable eigenvector for $\varepsilon=0.3$
and $\nu=0.01$.
The eigenvector has the shift-and-reflect symmetry
$\phi(x+\pi,-y) = \phi(x,y)$
and has the same periodicity in $x$ as~$\psi_M$. The
growth rates of the two different modes cross at around the
value $\nu=0.5$, as can be seen in Fig.~\ref{2valpour0.3}.

With the use of a simplified heuristic model it is possible to
understand the two modes. Since the flow is mainly
present in a confined region, then, for the sake of
simplicity, let us consider the main flow to consist of
counter-rotating vortices in a finite box: $\psi_0=A\sin(kx)
\cos(\ell y)$. We approximate the marginally unstable
shearing mode by $\phi_1=B\cos(\ell y)$, as suggested by
Fig.~\ref{Eigenvector}(a). With this ansatz, one can then
show that the Jacobian term in Eq.~(\ref{eq:equationenphi})
will give
\begin{eqnarray}
J\left( \psi_0,\nabla^2\phi_1\right)
+J\left(\phi_1,\nabla^2\psi_0\right)
=-{AB\ell k^3\over2} \cos(kx)\sin(2\ell y)\quad .
\end{eqnarray}

The interaction of the basic flow $\psi_0$
with the marginally unstable mode $\phi_1$
thus generates
a third term  $\phi_2=C\cos(kx)\sin(2\ell y)$,
completing the triad. We therefore consider
an unstable perturbative mode of the form
$(\phi_1+\phi_2)$, which will have the same pattern as the
most unstable mode found  for $\nu=0.01$ and shown in
Fig~\ref{Eigenvector}(b).

We can go further and explain the occurrence of the
instability. We continue to approximate
the base flow by $\psi_0=A\sin(kx)
\cos(\ell y)$, with $A$ fixed, and the perturbation by
\begin{eqnarray}
\phi_1+\phi_2=B(t)\cos(\ell y)+C(t)\cos(kx)\sin(2\ell y)\quad .
\end{eqnarray}
We substitute these approximations into
Eq.~(\ref{eq:equationenphi}):
\begin{equation}
{\partial  \nabla^2(\phi_1+\phi_2)\over \partial t}=
J(\psi_0, \nabla^2(\phi_1+\phi_2))+
J((\phi_1+\phi_2),\nabla^2\psi_0)
+\nu\nabla^4(\phi_1+\phi_2)\quad ,
\label{eq:equationenphimodel}
\end{equation}
obtaining
\begin{eqnarray}
-\dot B\ell^2\cos(\ell y)&-&(k^2+4\ell^2)\dot
C\cos(kx)\sin(2\ell y)=-{ABk^3\ell\over 2}\cos(kx)\sin(2\ell
y)\nonumber\\ &-&{3\over 4}k\ell^3 A\ C\left[\cos(\ell y)
+3\cos(2kx)\cos(\ell y)+3\cos(3\ell y)+\cos(2kx)\cos(3\ell
y)\right]\nonumber\\ &+&\nu \left(\ell^4\ B\cos(\ell
y)+(k^2+4\ell^2)^2\ C\cos(kx)\sin(2\ell y)\right)\quad.
\label{grossegalerkin2}
\end{eqnarray}

Projecting onto $\cos(\ell y)$ and $\cos(kx)\sin(2\ell y)$
gives the following
Galerkin system for the time dependent amplitudes $B$ and $C$:
\begin{mathletters}
\begin{eqnarray}
\dot B=&-\nu \ell^2\ B&-{3k\over 4}\ell A\ C
\label{galerkin1}\\ \dot C=&-\nu(k^2+4\ell^2)\ C&-{k^3\ell
\over2(k^2+4\ell^2)}A\ B\label{galerkin2} \end{eqnarray}
\end{mathletters}
Finally, looking for solutions $B=B_0\ e^{st}$
and $C=C_0\ e^{st}$,
one gets the equation
\begin{eqnarray}
s^2+s\nu(k^2+5\ell^2)+\nu^2\ell^2(k^2+4\ell^2)-
{3k^4\ell^2A^2\over 8(k^2+4\ell^2)}=0\quad .
\label{equationdudeuxiemedegree}
\end{eqnarray}

For low values of $A$ (i.e. low value of the intensity of the
electric  current: see section~\ref{Discussion}),
the initial flow is stable since all
solutions of equation~(\ref{equationdudeuxiemedegree}) are negative.
In contrast, above the threshold value
\begin{eqnarray}
A_c={2\nu(k^2+4\ell^2)\over k^2}\sqrt{2\over3}\quad,
\end{eqnarray}
one solution of Eq.~(\ref{equationdudeuxiemedegree})
is real and positive:
we get a stationary bifurcation giving rise to an
instability of the perturbation  $(\phi_1+\phi_2)$
whose pattern coincides with that shown in
Fig.~\ref{Eigenvector}(b). This simple approach therefore gives
a good qualitative understanding  of the relationship
between the marginal and unstable modes, and of the onset of
instability.

Figure~\ref{valpour0.20.30.5} shows the evolution of the
greatest eigenvalue versus viscosity  for three different
values of $\varepsilon$ on a logarithmic scale.
We see that the results are independent of $\varepsilon$
in the diffusive regime (high viscosity), as explained in
the preceding section. The inset allows us to ascertain that
the value of $\lambda$ is a linear decreasing function of
the viscosity. In the low viscosity regime the evolution is
qualitatively the same but the curves are distinct. The
bigger the parameter $\varepsilon$, the bigger the leading
eigenvalue and,  therefore the more unstable the flow. One
can  also  verify that the Mallier-Maslowe vortices with
$\varepsilon=0.5$ become unstable at a critical
 viscosity~$\nu_c$ which is slightly higher than that
corresponding to $\varepsilon=0.2$:  the critical
viscosity~$\nu_c$ is an increasing function of $\varepsilon$.

\subsection{Relation to the experiment}
\label{Discussion}

In the experiment by Tabeling et al.~\cite{TABELINGPERRINFAUVE},
the typical velocity $V$ of the basic regime can be
found by balancing the
forcing with the  viscous term,
as we have done via our assumption~(\ref{forceext}).
In dimensional terms, this leads to
the relation \begin{equation} V={BhI\over  \ell\nu \rho}
\end{equation} in which $B$ is the maximum value of the
magnetic field, $I$ the intensity of the electric current,
$h$ the depth of the fluid layer, $\ell$ the width of the
magnet, $\nu$ the viscosity and $\rho$ the density of the
fluid. Because the typical velocity varies linearly
with $I$, it is reasonable to suppose that the streamfunction
and its derivatives will also increase linearly with $I$.
We thus take our streamfunction $\psi=\xi\psi_M$,
where $\psi_M$ is the Mallier-Maslowe
solution~(\ref{Stuart+mallier})
and $\xi$ is a scalar which increases with $I$.

One can easily check that if $\psi_e$ is any solution to
the stationary solution Euler
equation~(\ref{eq:equationenpsi}), then  $\xi\psi_e$ is
also a solution. Let us show that the linear stability of all the
solutions $\xi\psi_e$ is determined by a linear stability
analysis of $\psi_e$ as a function of viscosity. The
equation governing the evolution of an infinitesimal
perturbation $\phi$  to the new inviscid solution is
\begin{equation}
{\partial  \nabla^2\phi\over \partial t}=J(\xi\psi_e, \nabla^2\phi)+
J(\phi,\nabla^2\xi\psi_e)+\nu\nabla^4\phi\quad .
\label{eq:equationenphiprim}
\end{equation}
Dividing by $\xi$, we get
\begin{equation}
{1\over \xi}{\partial  \nabla^2\phi\over \partial t}=J(\psi_e,
\nabla^2\phi)+ J(\phi,\nabla^2\psi_e)+{\nu\over
\xi}\nabla^4\phi\quad .
\label{eq:eqstablin}
\end{equation}
Thus, studying  the linear stability of the
solution $\xi\psi_e$ for viscosity $\nu$ is equivalent to
studying  the linear stability of the solution $\psi_e$ for
 viscosity $(\nu/\xi)$, except that the eigenvalue will also be
modified by $\xi$.

Using this insight, it is then possible to understand the
appearance of the instability. Recall from
Fig.~\ref{2valpour0.3} that the counter-rotating vortex
flow $\psi_M$ is stable for sufficiently high~$\nu$. Thus
$\xi\psi_M$ is stable for sufficiently high ${\nu/
\xi}$, i.e., for sufficiently low electric current.
Increasing the electric current $I$ in the experiment
corresponds to increasing $\xi$ and,  therefore, to decreasing
the ``renormalized'' viscosity $(\nu/\xi)$. The solution
will therefore remain stable until the renormalized
viscosity reaches the critical viscosity $\nu_c$ shown in
Fig.~\ref{2valpour0.3}. Above this threshold, the
counter-rotating vortices will be unstable and will evolve
as presented in the next section. The appearance of the
instability of the counter-rotating vortices for a high
enough electric current $I$ is thus explained.

\section{TIME INTEGRATION AND CONCLUSION}
\label{timeintegration}

The transitions resulting from the linear instability of the
Mallier-Maslowe vortices are studied by time-integrating the
Navier-Stokes equation~(\ref{eq:equationdeNScomplete}).
This numerical
experiment is constructed to
resemble that of Tabeling et
al, except that the size of the box is infinite.
  As an initial condition,
we add to the Mallier-Maslowe vortex flow~$\psi_M$
 a small perturbation
of the form $\phi(x,y,t=0)=\exp(-y^2)\cos x$, to accelerate
the appearance
of the possible unstable modes. The spatial representation
is, as discussed in section~\ref{Numericalprocedure}, with
a resolution of $N_x=32$, $N_y=69$  and $y_{\max}=15$. The
time stepping is carried out according to
Eq.~(\ref{addi}), with $\Delta t=0.01$.

If the simulation is carried out with $\varepsilon=0.3$ and
$\nu=5$, the patterns are stable, confirming the
linear stability analysis presented in the previous section.
When the parameters $\varepsilon$ and $\nu$ are
fixed at 0.3 and 0.01 respectively,
 the evolution, depicted in Fig.~(\ref{evolutionpournu}),
 is clearly different. We see in
Fig.~\ref{evolutionpournu}(b) that at $t=65$ we have a
linear array of tilted vortices of positive sign and the
size of the vortices has doubled as occurred in the
experiment. The negative vortices have been ejected away
from the center of the box.

If we plot the deviation $\phi=(\psi-\psi_M)$ from
 the initial condition $\psi_M$, we find in
the initial stage (at $t=50$ for example)
the streamfunction presented in figure~\ref{evolutionsanspsi}(a),
confirming the linear stability analysis (see
Fig.~\ref{Eigenvector}(b)). After this linear transient
growth, the flow continues to evolve until it approaches the
pattern depicted in figure~\ref{evolutionpournu}(c).
However, the amplitude continues to increase with time (see
Fig.~\ref{normedeuxcas}), while preserving the pattern.
Figure~\ref{evolutionsanspsi}(b) depicts the deviation
$\phi$  at $t=200$
and we note that  the pattern around the $x$-axis
resembles somewhat the Kelvin-Stuart vortices  (see
Fig.~\ref{profileofthesolution}(b)). This explains why the
system does not reach a final equilibrium state:
Since~$\phi$  is then itself a
solution of the stationary inviscid Eq.~(\ref{stationaireEuler})
$J(\phi,\nabla^2\phi)=0$, the nonlinear
saturation effect disappears and $\phi$ continues to
grow in time. A similar case was found
in zero Prandtl number convection, where the linearly
unstable roll modes are exact nonlinear solutions~\cite{KUMARFAUVETHUAL}.

One possible reason that our simulation,
unlike the experiment of
Tabeling et al~\cite{TABELINGPERRINFAUVE},
does not reach a final equilibrium state
could be that we considered the flow to be
perfectly two-dimensional. In the experiment,
two-dimensionality
is enforced by using a shallow fluid layer: a
frictional force proportional to the velocity could
capture this bottom-friction
effect~\cite{GOTOH,THESS}. The addition of a term
$-\mu\nabla^2\psi$
proportional to the velocity on the right hand side of the
equation~(\ref{eq:equationdeNScomplete}) will change the
evolution of the flow. The eigenvalues in the presence of
linear friction differ from those of the problem without
friction only by a shift $(-\mu)$ of the growth rate: thus
the determination of the dependence of the eigenvalue
spectrum on $\mu$ does not require additional numerical
studies and, in addition, the linear friction is always
stabilizing.

We therefore time-integrate  this system,
including the linear friction term as well as the
ordinary viscosity. The external
force is now chosen as
\begin{equation}
G_{ext}=-\nu\nabla^4\psi_M+\mu\nabla^2\psi_M
\end{equation}
so that, as in the previous section,
the viscosity only acts on the perturbation not the
basic flow.
The resulting streamfunction at  $t=300$ is shown in
Fig.~\ref{evolutionsanspsimaislammbda} for a small value
of $\mu$ fixed at 0.01. Contrary to the
case without linear friction ($\mu=0$, see
Fig.~\ref{evolutionpournu}), the flow attains an equilibrium
state with co-rotating vortices along the $x$-axis as
demonstrated by the time series in
 Fig.~\ref{normedeuxcas}. Thus, the
linear friction term stabilizes  the row of co-rotating
vortices as was obtained in the experiment. The necessity of
this linear friction  term in reproducing the final state of
the experiment could be a reason why the final maximum
entropy state is not often reached by Marteau and
al~\cite{MARTEAUCATTAB} in their experiment: in their small
2D lattice of electromagnetically forced vortices, the
bottom-friction effect should also be important.

The purpose of this work was to understand and explain the
behavior of the instructive experiment of
Tabeling et al.~\cite{TABELINGPERRINFAUVE}. First, we derived
explicitly the  nonlinear stability condition for the
counter-rotating vortex solutions in a rectangular box
without compensating viscosity; the finite-size effects of the box
were studied. Then, introducing viscosity and
compensating forcing, we
derived a model for the appearance of the instability when the
electric current is increased: the renormalized viscosity
decreases until it reaches the  critical viscosity $\nu_c$
determined by a numerical linear stability analysis.
 We found $\nu_c$  to be
an increasing function of the parameter $\varepsilon$
characterizing the vorticity which could explain the fact
low vorticity-flows (i.e. low $\varepsilon$ in
Eq.~(\ref{sinhPoisson})) are more visible in 2D
hydrodynamic flows, since they are more stable than
high vorticity-flows.
Above this threshold, the evolution of the system leads to a
final equilibrium state similar to that in the experiment
if a linear term is added to the standard
Navier-Stokes equation to reproduce
the effects of the friction at the bottom of the experimental
apparatus.

\acknowledgements
The authors thank D. D. Holm, A.C. Newell and S. Takeno
for helpful discussions.

\begin{figure}
\caption{Steady Flows.
Figure (a)  represents the streamlines of the
Mallier-Maslowe solution~({\protect
\ref{eq:solutionalternee}})  for $\varepsilon=0.3$.
Figure (b) represents the streamlines of the  Kelvin-Stuart
solution~({\protect\ref{equationdeStuart}}) for $\varepsilon=0.3$.
The dashed curves are negative contour lines and
the solid curves are positives ones. }
\label{profileofthesolution}
\bigskip\bigskip

\caption{The region A defines the  domain of sufficient
stability of a pair of counter-rotating vortices ($N=1$) in
the plane ($\ell,\varepsilon$). $\ell$~is the transverse
size of the box and $\varepsilon$ characterizes the
vorticity density. The solid line is defined by
equation~({\protect \ref{eq:conditionstab}}).}
\label{lfonctiondeepsilon}
\bigskip\bigskip

\caption{Eigenvalues. Dependence of the growth rate for the
first mode on the  viscosity $\nu$ for the
counter-rotating vortices when $\varepsilon=0.3$. The
squares and the dashed curve correspond to the most unstable
(or least stable) eigenvalue
and the asterisks and the solid curve to the $x$-independent mode.}
\label{2valpour0.3}
\bigskip\bigskip

\caption{Eigenvectors. The streamlines of the most
unstable eigenvector associated with the Mallier-Maslowe vortices
for $\varepsilon=0.3$ are presented as a
surface plot for $\nu=10$ in figure (a),
while figure (b) depicts the contour-plot when $\nu=0.01$.
The eigenvalues associated with these eigenvectors
are  $\lambda=-0.0115$ and $\lambda=0.099$, respectively.}
\label{Eigenvector}
\bigskip\bigskip

\caption{Dependence of the growth rate on viscosity for
three values of $\varepsilon$ for the counter-rotating
vortices. The diamonds and the dash-dotted curve correspond
to $\varepsilon =0.2$,  the squares and the
dash-triple-dotted curve correspond to $\varepsilon =0.3$
and the triangles and the dashed curve correspond to
$\varepsilon =0.5$. The asterisks and the solid curve represent the
$x$-independent mode. A logarithmic scale is used for the viscosity;
the inset uses a linear scale. }
\label{valpour0.20.30.5}
\bigskip\bigskip

\caption{Contour plot of the stream function for
$\varepsilon =0.3$ and $\nu=0.01$. (a) initial condition,
the Mallier-Maslowe~$\psi_M$; (b) $t=65$;
(c) $t=200$.}
\label{evolutionpournu}
\bigskip\bigskip

\caption{Contour plot of the deviation $\phi=\psi-\psi_M$
from the Mallier-Maslowe solution for $\varepsilon =0.3$,
$\nu=0.01$ and $\mu=0$. (a) $t=50$; (b) $t=3000$.}
\label{evolutionsanspsi}
\bigskip\bigskip

\caption{Evolution of the norm of the deviation
$||\phi||=||\psi-\psi_M||$  from the Mallier-Maslowe solution for
$\varepsilon =0.3$, $\nu=0.01$ versus time. The solid
curve corresponds to $\mu=0$ and the dashed curve to $\mu=0.01$.}
\label{normedeuxcas}
\bigskip\bigskip

\caption{Contour plot of the stream function $\psi$
for $\varepsilon =0.3$, $\nu=0.01$ and $\mu=0.01$ at
$t=5000$.}
\label{evolutionsanspsimaislammbda}
\end{figure}
\end{document}